\documentclass[prl,twocolumn,showpacs,superscriptaddress,amsmath,amssymb,aps]{revtex4-1}
\usepackage{amsmath}
\usepackage{bm}              
\usepackage{graphicx}

\begin{document}


\title{Optical properties of superconducting EuFe$_2$(As$_{1-x}$P$_x$)$_2$}

\author{David Neubauer}
	\email{david.neubauer@pi1.physik.uni-stuttgart.de}
		\affiliation{1.Physikalisches Institut, Universit\"at Stuttgart, Pfaffenwaldring 57, 70550 Stuttgart, Germany}
	
\author{Artem V. Pronin},
		\affiliation{1.Physikalisches Institut, Universit\"at Stuttgart, Pfaffenwaldring 57, 70550 Stuttgart, Germany}
	
 \author{Sina Zapf}
		\affiliation{1.Physikalisches Institut, Universit\"at Stuttgart, Pfaffenwaldring 57, 70550 Stuttgart, Germany}
	
  \author{Johannes Merz}
		\affiliation{1.Physikalisches Institut, Universit\"at Stuttgart, Pfaffenwaldring 57, 70550 Stuttgart, Germany}
		
    \author{Hirale S. Jeevan}
			\affiliation{I. Physikalisches Institut, Universit\"at G\"ottingen, Friedrich-Hund-Platz 1, 37077 G\"ottingen, Germany}
			\affiliation{Department of Physics, PESITM, Sagar Road, 577204 Shimoga, India}
			
    \author{Wen-He Jiao}
			\affiliation{Department of Physics, Zhejiang University, Hangzhou 310027, China}
			
    \author{Philipp Gegenwart}
			\affiliation{I. Physikalisches Institut, Universit\"at G\"ottingen, Friedrich-Hund-Platz 1, 37077 G\"ottingen, Germany}
			\affiliation{Experimentalphysik VI, Universit\"at Augsburg, Universit\"atsstra\ss e 1, 86135 Augsburg, Germany}
		
    \author{Guang-Han Cao}
			\affiliation{Department of Physics, Zhejiang University, Hangzhou 310027, China}
			
    \author{Martin Dressel}
			\affiliation{1.Physikalisches Institut, Universit\"at Stuttgart, Pfaffenwaldring 57, 70550 Stuttgart, Germany}

\begin{abstract}%

We present a broadband optical-conductivity study of superconducting
single-crystalline EuFe$_2$(As$_{1-x}$P$_x$)$_2$ with three
different substitutional levels. We analyze the normal-state electrodynamics
by decomposing the conductivity spectra using a
Drude-Lorentz model with two Drude terms representing two
groups of carriers with different scattering rates. The analysis
reveals that the scattering rate of at least one of the Drude
components develops linearly with temperature for each doping level.
This points towards strong electron-electron correlations and a
non-Fermi-liquid behavior in the P-substituted superconducting Eu-122 pnictides.
We also detect a transfer of the spectral weight from mid-infrared
to higher frequencies and assign it to the Hund's-rule coupling
between itinerant and localized carriers. The conductivity spectra below the superconducting transition show no sharp features
to be associated with the dirty-limit superconducting BCS gaps. We interpret these
results in terms of clean-limit superconductivity in
EuFe$_2$(As$_{1-x}$P$_x$)$_2$. The best parametrization fit can be
achieved using a two-gap model. We find that the larger gap at the
hole pockets of the Fermi surface is likely to be isotropic, while
the smaller gap at the electron pockets is anisotropic or even
nodal.

\end{abstract}
\maketitle  
\section{Introduction}

Among the superconducting iron pnictides, the
so-called 122-family ($A$Fe$_2$As$_2$ with $A$=Ba, Sr, Ca, Eu) is
the most studied composition. The superconductivity in this family
can be achieved not only by doping the antiferromagnetic
parent compounds with charge carriers, but also by pressure, either external
or chemical \cite{Miclea2008,Ren2009}. Substituting As by isoelectric P is
often considered as a ``clean" route towards superconductivity
\cite{VanderBeek2010} because in this case the conducting Fe layers
remain undisturbed, the quasiparticle scattering is presumably not
increased by the presence of dopants, and hence the inherent
superconducting properties manifest themselves in the most vivid
way. Indeed, most convincing experimental evidence for such
remarkable properties as the presence of a quantum critical point
(QCP) beneath the superconducting dome
\cite{Shibauchi2014,Hashimoto2012} or the existence of the line
nodes in the superconducting gap \cite{Hashimoto2010} have been
obtained on P-substituted BaFe$_2$As$_2$.

Among the 122-family, the Eu-based systems are of particular
interest because of their unique properties originating from the
strong local-moment magnetism of the Eu$^{2+}$ ions
\cite{Jiang2009,Zapf2011,Zapf2013,Zapf2014,Nandi2014a,Cao2011,Jeevan2011,KrugvonNidda2012}.
More investigations on EuFe$_2$(As$_{1-x}$P$_x$)$_2$ are currently
highly desirable to further examine the impact of the Eu$^{2+}$
magnetism on electronic properties and to eventually compare the
Eu-based systems with other members of the $A$Fe$_2$As$_2$ family in
this regard. In order to fulfill this task, we performed an optical
investigation of EuFe$_2$(As$_{1-x}$P$_x$)$_2$. Optical spectroscopy
is a proven powerful tool to gain insight into the electrodynamics
of correlated-electron materials in general \cite{Basov2011} and of
iron pnictides in particular \cite{Charnukha2014}.

\section{Experiment}

We investigated EuFe$_2$(As$_{1-x}$P$_x$)$_2$ single crystals with
three different levels of phosphorus content ($x=0.15$, $0.165$, and
$0.21$). The samples with $x=0.15$ and $0.21$ (P015, P021) were
grown at Zhejiang University in Hangzhou, China, via spontaneous nucleations, similar to the previous report \cite{Jiang2009}, but using Fe(As$_{1-x}$P$_x$) as the self flux. 
The $x=0.165$ sample (P0165) was synthesized at G\"ottingen
University in Germany utilizing the Bridgeman method \cite{Jeevan2011}. The chemical compositions were found by 
energy-dispersive x-ray spectroscopy.

The dc resistivity measurements were made in a standard four-probe
geometry using a home-built cryogenic setup. The magnetization
measurements were performed in a Quantum Design MPMS $7$ T system in
the static mode. All measurements presented in this article,
including resistivity, magnetization, and optical reflectivity, were
performed for the \textit{ab} plane of EuFe$_2$(As$_{1-x}$P$_x$)$_2$
crystals (the ``in-plane" geometry).

The optical reflectivity, $R(\omega)$, was investigated for
temperatures between $3.5$ and $300$\,K over a wide frequency range
from $12$ to $12\,000$~cm$^{-1}$. Additionally, reflectivity
measurements up to $22\,000$~cm$^{-1}$ were performed on sample P015 at room-temperature. Measurements at
far-infrared frequencies (FIR, $12 - 1000$~cm$^{-1}$) were carried
out with a Bruker IFS 113v Fourier-transform infrared (FTIR)
spectrometer utilizing several beam-splitters and bolometric
detectors operating at 4.2 and 1.3\,K. For referencing, \textit{in
situ} gold coating of the samples was performed in this frequency
range. At higher frequencies (i.e. in the mid- and near-infrared as
well as in the visible ranges; MIR, NIR, VIS; $800 -
22\,000$~cm$^{-1}$) the measurements were conducted using a Bruker
Hyperion infrared microscope attached to a Bruker Vertex 80v FTIR
spectrometer. Freshly evaporated gold mirrors acted as references in
MIR and NIR, protected silver mirrors were used in the VIS range.

The Kramers-Kronig analysis of the reflectivity spectra was
performed in two stages. For the initial transformation, the
measured $R(\omega)$ spectra were extrapolated to infinite frequency
using first the room-temperature measurements on P015 up to
$22\,000$~cm$^{-1}$ and then the standard free-electron behavior,
$R(\omega)\propto\omega^{-4}$. Using the room-temperature data from
one of the samples at these high frequencies is justified
by the fact that $R(\omega)$ has neither temperature nor
doping-level dependence at $\omega \geq 10\,000$~cm$^{-1}$, see
Fig.~\ref{fig:ref-sig}. At this stage, the zero-frequency
extrapolations were made using the Hagen-Rubens relation,
$[1-R(\omega)] \propto \sqrt{\omega}$, in the normal (metallic)
state and $[1-R(\omega)] \propto \omega^2$ in the superconducting
state. The optical conductivity spectra,
$\sigma(\omega)=\sigma_{1}(\omega)+i\sigma_{2}(\omega)$, obtained
from this initial Kramers-Kronig transformation were subsequently
fitted together with the measured $R(\omega)$ using the
Drude-Lorentz model discussed below. For the final Kramers-Kronig
transformation, the free-electron behavior at $\omega \rightarrow
\infty$ and the low-frequency extrapolations of $R(\omega)$ were
replaced by the results obtained from this fit. The experimental
data below 200\,cm$^{-1}$ were moderately smoothed prior to the
initial transformation to avoid excessive noise in the optical
conductivity at low frequencies.

\section{Resistivity and magnetic susceptibility}

\begin{figure*}[ht]
  \includegraphics*[width=\textwidth]{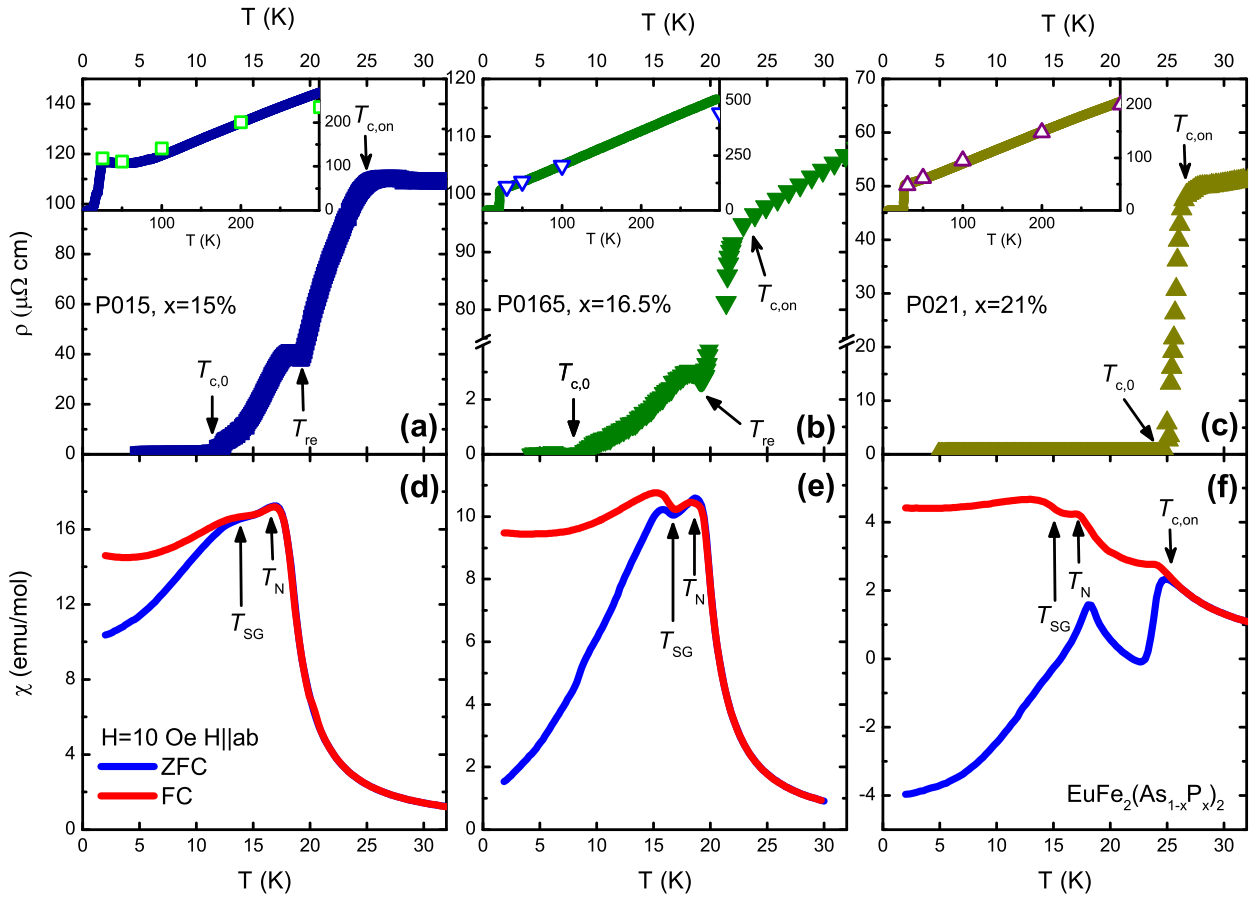}
\caption[]{\textbf{Top panels:} Temperature-dependent in-plane dc
resistivity, $\rho(T)$, of EuFe$_2$(As$_{1-x}$P$_x$)$_2$ for x=15\%
(a), x=16.5\% (b) and x=21\% (c). Main panels show $\rho(T)$ near
the superconducting transitions, while the complete resistivity
curves up to 300 K are given in the insets. Arrows at
$T_\text{c,on}$ mark onsets of the superconducting transitions;
$T_\text{re}$ refers to the resistivity re-entrance temperatures in
samples P015 and P0165; $T_\text{c,0}$ indicates the temperatures,
where zero resistivity is reached. Open symbols in the insets
correspond to the zero-frequency limits of the inverse optical
conductivity, $1/\sigma_1(\omega \rightarrow 0)$, obtained as
described in the text. \textbf{Bottom panels:} Molar in-plane
susceptibility, $\chi_\text{mol}(T)$, for the same samples, as
indicated in the corresponding top panels. At $T_\text{N}$, magnetic ordering of the Eu$^{2+}$ moments takes place,
while spin-glass transitions occur at $T_\text{SG}$ \cite{Zapf2013}.
Both types of magnetic transitions are seen in the zero-field cooled
(ZFC, blue curves) as well as in the field-cooled (FC, red curves)
measurements, the spin-glass transitions being most pronounced in
the latter. Note different vertical scales in the panels and the broken vertical scale in panel (b).}
    \label{fig:res-mag}
\end{figure*}

Fig.~\ref{fig:res-mag} shows temperature variation of the in-plane
resistivity $\rho$ and the molar magnetic susceptibility
$\chi_\text{mol}$ for the samples investigated. At high
temperatures, all three samples exhibit resistivity linear in
temperature. For samples P0165 and P021, this linearity persists
down to the onset of superconductivity. For P015, however, the
resistivity levels off below $T\approx$75\,K and stays constant
until superconductivity sets in. Such behavior of resistivity in
P015 is reminiscent of resistivity in strongly underdoped
EuFe$_2$(As$_{1-x}$P$_x$)$_2$, where the spin-density-wave (SDW)
ordering of the Fe$^{2+}$ moments is responsible for the leveling
off the $\rho (T)$ curve \cite{Jeevan2011}. According to Ref. \cite{Jeevan2011}, the doping
level of the crystal and the temperature, where the leveling-off
happens, appear to be too high to make the SDW-formation scenario
probable. A detailed transport study of EuFe$_2$(As$_{1-x}$P$_x$)$_2$ under hydrostatic pressure \cite{Tokiwa2012} puts the P015 sample in a range of the phasediagram where the SDW transition still appears in the temperature range where we observe the anomaly. However, in contrast to Ref. \cite{Tokiwa2012} we cannot identify any discontinuity in the resistivity of P015, that could be assigned to a SDW transition.

As one can see from the panels (a-c) of Fig.~\ref{fig:res-mag}, the
three samples show onsets of superconductivity at
$T_\text{c,on}\approx 25$\,K, $22$\,K, and $27$\,K, respectively.

Both specimens, P015 and P0165, have a rather broad superconducting transition
affected by a pronounced resistivity re-entrance around $19$\,K.
This re-entrant resistivity is a well-known phenomenon for Eu-based
122 iron pnictides; it stems from ordering of the Eu$^{2+}$ moments
at $T_\text{re}\approx 19$\,K, which has no significant doping
dependence  for underdoped and superconducting samples \cite{Zapf2013,Jiao2012,Cao2012}. The P021 sample
exhibits the highest $T_\text{c,on}$ and reaches zero resistance at
$T_\text{c,0}\approx$ 25\,K, i.e. well above the Eu-ordering
temperature. Thus, superconductivity is fully developed at 19\,K and
no re-entrance of resistivity is observed. A summary on the
transition temperatures, resistivity values at 300\,K and the
residual resistivity ratios (RRR) for the investigated samples is
given in Table \ref{table}.

Panels (d-f) of Fig.~\ref{fig:res-mag} display the results of the
in-plane magnetization measurements, performed in the static (dc)
mode. All three samples exhibit a complex magnetic behavior, which
is typical for the Eu-122 family
\cite{Zapf2011,Zapf2013,Jeevan2011}. In the case of samples P015 and
P0165, two clear humps can be resolved in the zero-field-cooled
(ZFC) and field-cooled (FC) curves. It has been shown, that the
feature at $T_\text{N}$ can be assigned to the magnetic ordering of the Eu$^{2+}$ moments with
strong ferromagnetic component out of plane
\cite{Nandi2014,Zapf2011,Zapf2013}. The dip at $T_\text{SG}$ preceding the
second hump is identified as a transition to an in-plane glass-like
behavior of the Eu spins \cite{Zapf2013}.

Although according to Figs.~\ref{fig:res-mag} (d-f) a negative in-plane signal, typical for diamagnetism, 
is only seen in P021, we are confident that bulk
superconductivity develops in all three samples. Measurements of
the static magnetic susceptibility with the field applied
perpendicular to the \textit{ab}-plane (i.e. when the screening
currents flow within this plane) do show a strong diamagnetic signal
also in the samples P015 and P0165 (not displayed). Furthermore, it was demonstrated recently that P0165 exhibits a strong diamagnetic signal in the measurements
of the in-plane \textit{dynamic} magnetic susceptibility with a
large constant magnetic field applied to suppress the ordering of
the Eu$^{2+}$ moments \cite{Zapf2013}.

\section{Optics: the normal state}

\begin{figure*}[ht]

 \includegraphics*[width=\textwidth]{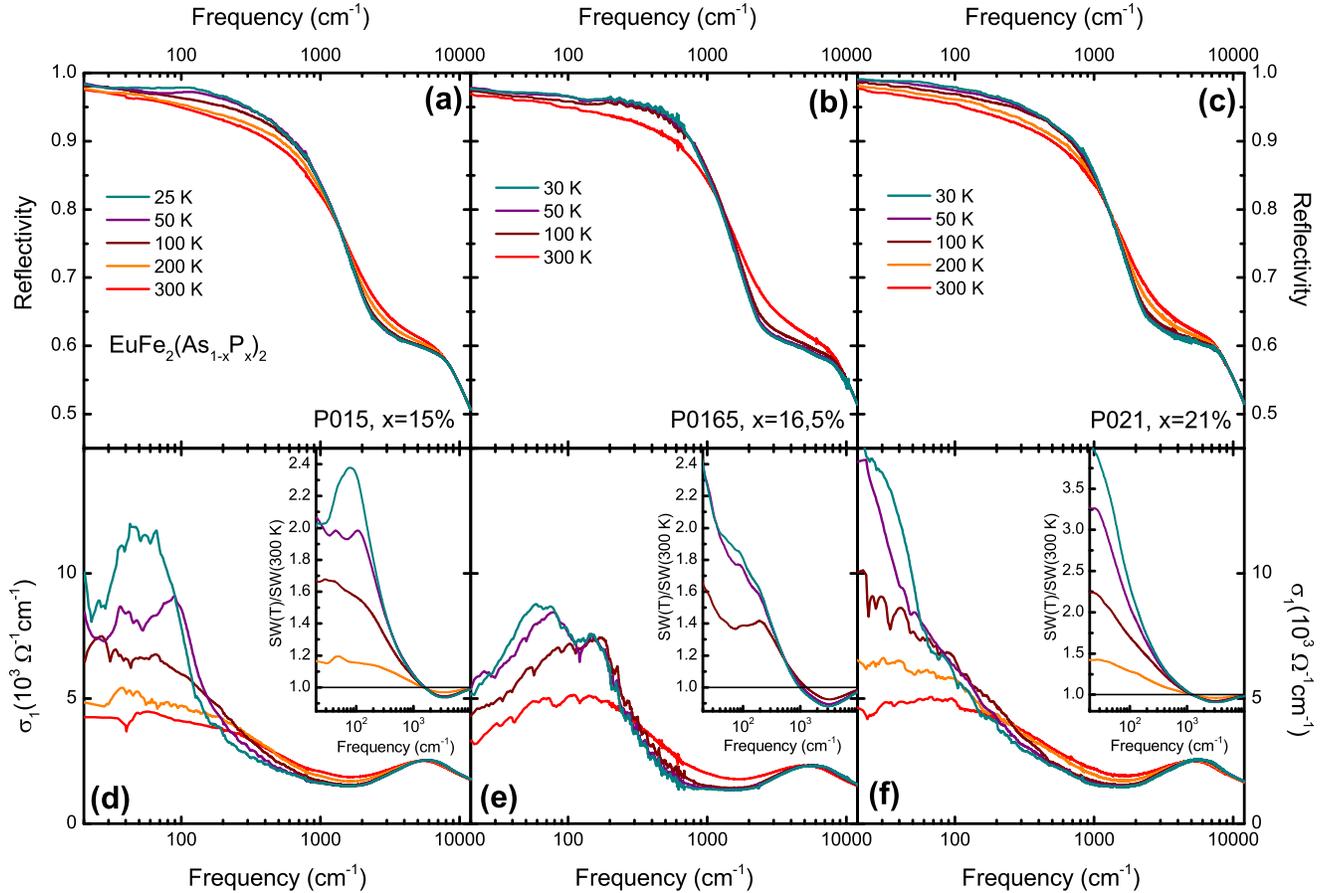}
 \caption[]{Frequency-dependent in-plane reflectivity of
EuFe$_2$(As$_{1-x}$P$_x$)$_2$ for $x=15$\% (a), $x=16.5$\% (b), and
$x=21$\% (c) at temperatures $T=$ $300$\,K (red), $200$\,K (orange),
$100$\,K (brown), $50$\,K (purple) and at (or slightly above)
$T_\text{c,on}$ (cyan; $25$\,K for $x=15$\%, $30$\,K for the other two
doping levels). Panels (d-f) display the real part of the optical
conductivity obtained by Kramers-Kronig transformation for $x=15$\%,
$16.5$\%, and $21$\%, respectively. The development of localization
modes in the FIR is clearly visible in the conductivity spectra for
$x=15$\% and $x=16.5$\%. The normalized spectral weight,
SW(T)/SW($300$\,K), displayed in the insets of panels (d-f),
uncovers a spectral-weight transfer to higher energies with
decreasing temperature.}
    \label{fig:ref-sig}
\end{figure*}

Panels (a-c) of Fig.~\ref{fig:ref-sig} display the normal-state
in-plane reflectivity of the investigated samples. All three
compounds demonstrate a qualitatively similar temperature evolution of
$R(\omega)$. The reflectivity in the FIR range increases with
lowering temperature, indicating the metallic nature of our
samples. Oppositely, the reflectivity at MIR frequencies ($\sim
2\,000 - 10\,000$ cm$^{-1}$) decreases as temperature goes down.
This behavior is usually ascribed to electron correlations arising
from the Hund's-rule coupling of electrons
\cite{Haule2009,Medici2011}. For all three crystals, there is a
narrow region of frequencies (around $1\,300$\,cm$^{-1}$) where
$R(\omega)$ is nearly temperature independent. Apart from these
common features, there is one important difference between the
reflectivity spectra of the samples:
while the FIR reflectivity is monotonic in frequency at any
temperature for P021, the $R(\omega)$ curves of the two other
samples exhibit a non-monotonic frequency behavior at $T \leq
100$ K for P0165 and at $T \leq 50$ for P015.

The real part of the optical conductivity, $\sigma_{1}(\omega)$,
obtained via the Kramers-Kronig transformation, is shown in the
bottom panels of Fig.~\ref{fig:ref-sig}. The non-monotonic FIR
reflectivity in P015 and P0165 manifest itself in $\sigma_1(\omega)$
as very broad modes dominating the FIR conductivity. As one can see
from Fig.~\ref{fig:ref-sig}, the development of the FIR mode has no
systematic doping dependence. Thus, we can conclude that the
appearance of the modes is a sample-related issue. In case of P015 we can speculate wether the FIR anomaly below $T = 100$\,K is due to a possible SDW transition discussed for this sample in the previous section. However, comparing the $\sigma_1(\omega)$ spectra to previous optical studies on electron or hole underdoped BaFe$_2$As$_2$ that clearly show a SDW transition, we notice that below T$_\text{SDW}$ a common feature is a pile up of spectral weight (SW) at approximately $1\,000$ cm$^{-1}$ while at lower frequencies the spectra get suppressed, irrespective of doping type or level \cite{Nakajima2010,Nakajima2012,Dai2012}. Hence, the mismatch in energy scale of our FIR feature together with the absence of any sharp feature in the resistivity in Fig. \ref{fig:res-mag} (a) enables us to conclude, that this FIR peculiarity in P015 is not related to a SDW ordering. In case of the P0165 sample, the same argument holds. We want to point out that the observation of similar modes in the FIR in iron pnictides is by far not exceptional. There are reports of FIR modes in hole-doped \cite{Kwon2012,Dai2012}, and electron-doped Ba-$122$ \cite{Lobo2010,VanHeumen2009}, even in the parent compound \cite{Charnukha2013}. For all these reports we can not see a systematic doping level dependence and there are also reports where any comparable feature is absent (e.g. Ref.'s \cite{Barisic2010,Dai2013a,Nakajima2010}). Even though the interpretation of these modes range from pseudogap to low lying interband transitions, we find, that due to the non-systematic appearance, localization of itinerant carriers induced by impurities \cite{Anderson1958} is the most probable origin of the modes, as already suggested in Refs. \cite{Lobo2010,Charnukha2013,Cheng2011}.

Typical metallic behavior of the optical conductivity, i.e. a
monotonic narrowing of the zero-frequency-centered Drude-like mode
with decreasing temperature, is clearly visible only for sample
P021. In P0165 and P015, the temperature development of the Drude
response is perturbed by the localization modes. Nevertheless, the
zero-frequency limit of the optical conductivity, 
$\sigma_1(\omega\rightarrow 0)$, nicely follows the dc data for all
the samples at any temperature, as it can be seen from the inserts
of Figs.~\ref{fig:res-mag} (a-c).

We now calculate the frequency-dependent spectral weight defined
as:
\begin{equation}
    SW(\omega)=\int^{\omega}_{0}\sigma_1(\omega') d\omega'.
        \label{eq:sw}
\end{equation}
For temperatures in the normal state, the spectral weight normalized
to the spectral weight at $T=300$\,K is displayed in the insets of
Figs.~\ref{fig:ref-sig} (d-f). The general trends of the
spectral-weight behavior are the same for all three samples,
although the localization modes affect the normalized spectral
weight at low frequencies, where it stays above 1 and is
maximal at the lowest temperatures. In the MIR range, the
temperature-dependent spectral weight is below its room-temperature
values; the lower the temperature, the more pronounced the losses are.
After passing a minimum at $3\,000 - 4\,000$\,cm$^{-1}$, the
normalized spectral weight increases again. Only at frequencies
above $20\,000$\,cm$^{-1}$ the spectral weight becomes eventually
conserved (within an uncertainty of $< 1$\%) for all the samples.
The presence of the spectral-weight transfer up to such high
energies is an important feature of iron
pnictides and is ascribed to correlation effects, e.g. to
the Hund's rule correlations mentioned above
\cite{Barisic2010,Wang2011,Schafgans2012}. In this regard, the
members of the Eu-122 family, investigated in this work, behave very
similarly to many other pnictides.

In order to get insight in the mechanisms contributing to the
frequency-dependent optical response, one needs to decompose the
measured spectra into a number of simple terms. We fitted our
experimental spectra in a standard way -- by a sum of Drude and
Lorentz terms. For the complex optical conductivity, the model reads
as \cite{Dressel2002}:
\begin{equation}
\sigma(\omega)=\sum_\text{Drude}\frac{\omega_{p}^{2}}{4\pi}\frac{1}
{1/\tau-i\omega}+\sum_\text{Lorentz}\frac{\omega_{p}^{2}}{4\pi}\frac{\omega}
{i(\omega_{0}^2-\omega^{2})+\omega/\tau}.\rm
\label{eq:dru-lor}
\end{equation}
Here $\omega_p$ is the plasma frequency of each term, $\omega_0$ is
the resonance frequency of each Lorentz term, and $1/\tau$ denotes
the scattering rates of free carriers in the case of Drude terms and
introduces a damping for each Lorentz oscillator.

Because of the multiband nature of iron pnictides, there are many
ways to decompose their optical spectra, as it has been already
discussed e.g. in Ref.~\cite{Zapf2015}. We use a model that captures
all major spectral features, but still keeps the number of free
parameters as small as possible. It has been shown that a
Drude-Lorentz model with two Drude components provides an adequate
description for the optical response of many iron pnictides
\cite{Wu2010b,Wu2009}. For the 122 family, it is well established,
that one of the two Drude components is rather broad in frequency
(large $1/\tau$) and the other one is quite narrow (small $1/\tau$).
Sometimes, the broad component is replaced by an overdamped
Lorentzian. The broad term is usually associated with the hole
pockets of the Fermi surface, while the narrow Drude component is
normally assigned to the electron pockets
\cite{Barisic2010,Zapf2015,Moon2010,Nakajima2010,Dai2013a}. This
assignment is based on the differences in scattering rates of the
two carrier types found in pnictides by Hall measurements: holes
posses a rather large $1/\tau$, while electrons scatter much less
frequently, i.e. their $1/\tau$ is smaller \cite{Fang2009}.
Thereafter, we abbreviate the broad and the narrow Drude components
as BD and ND, respectively. The fit parameters of the Drude
components for the lowest normal-state temperatures of all three
samples are given in Table~\ref{table}.

In addition to the two Drude terms, a number of Lorentzians are
necessary to fit the experimental spectra properly. We use a broad
Lorentzian (MIRL) to describe the
flat MIR background, which was previously reported in a number of
122 compounds and which is believed to be due to an
interband-transition band located at around $1\,000$\,cm$^{-1}$
\cite{Zapf2015,Marsik2013}. Another Lorentz term, marked as ``HUND", is used to fit a
pronounced broad interband-transition band located around
$6\,000$\,cm$^{-1}$. Both bands display a strong temperature dependence, which
is believed to stem from a strong renormalization by Hund's-rule
coupling effects \cite{Charnukha2014}. Additionally, we need to use
a few temperature-independent modes at higher frequencies. Let us note, that the optical conductivity was fitted together with
the measured reflectivity in order to provide the most consistent
description of the experimental spectra.

\begin{figure*}[ht]
\includegraphics*[width=\textwidth]{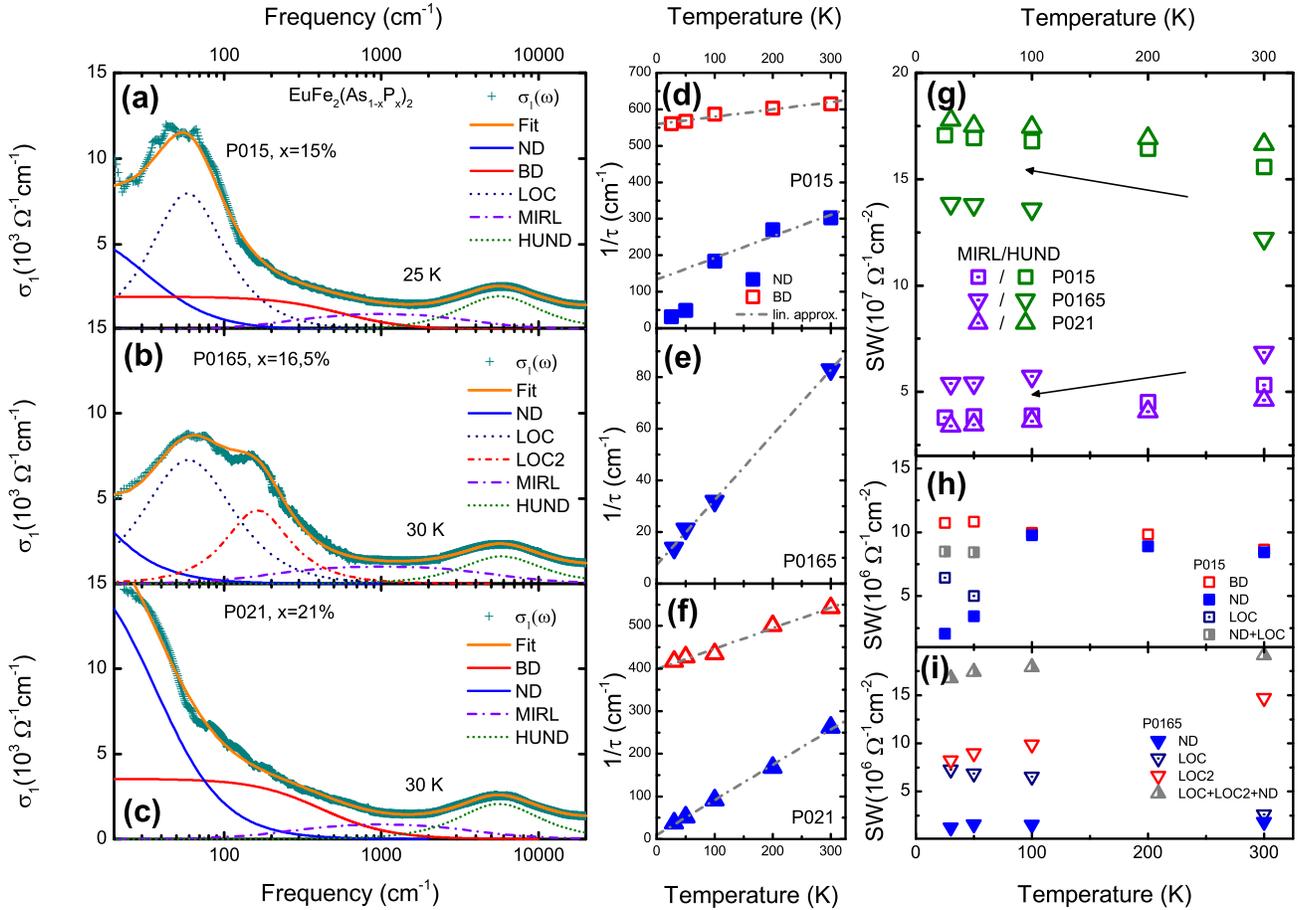}
\caption{\textbf{Left panels (a-c):} Frequency-dependent optical
conductivity of EuFe$_2$(As$_{1-x}$P$_x$)$_2$ for $x=15$\% (a),
$16.5$\% (b), and $21$\% (c) at the lowest measured normal-state
temperature (30\,K for $x=16.5$\% and $21$\%; $25$\,K for $x=15$\%).
Dark cyan crosses represent data points; orange lines are the
resultant Drude-Lorentz fits. Contributions of the single terms of
Eqn.~\ref{eq:dru-lor} are also displayed in the panels.
\textbf{Center panels (d-f):} Scattering rates of the narrow (ND)
and broad (BD) Drude terms for the three samples measured. Straight
lines are guides to the eye. \textbf{Panel (g):} Temperature
evolution of the spectral weights of the high-frequency MIRL and
HUND modes for all three sample. The spectral wight of the HUND mode
grows at the expense of the MIRL mode with decreasing temperature
for all three samples, as indicated with arrows. \textbf{Panels (h)
and (i):} Temperature evolution of the low-frequency single-term
spectral weights for samples P015 (h) and P0165 (i). In P021, the
spectral weights of the low-frequency (i.e. Drude) terms are
temperature independent.} \label{fig:fit}
\end{figure*}

For sample P021, the described model is sufficient for an adequate
description of the experiment [see Fig.~\ref{fig:fit} (c)]. For the
two other samples, we had to introduce more terms to describe
the localization bands. In the case of sample P015, one additional
Lorentzian (LOC) in the FIR range was sufficient [Fig.~\ref{fig:fit}
(a)]. For P0165, we had to introduce two Lorentzians (LOC and LOC2)
and to omit the BD term at all temperatures [Fig.~\ref{fig:fit}
(b)], as otherwise no reasonable description was possible \cite{Note1}. This
indicates that localization effects are particularly strong in
P0165.

We now discuss the temperature changes in the single terms of
Eqn.~\ref{eq:dru-lor}. First, we discuss the scattering rates
$1/\tau$ of the Drude terms shown in the middle panels of
Fig.~\ref{fig:fit}. In P021, the scattering rate of both, broad and
narrow, Drude terms is linear in temperature. The same behavior is
visible in the only (narrow) Drude component of P0165. This
observation together with the $T$-linear resistivity provides a
strong evidence for non-Fermi-liquid behavior which might be caused
by strong magnetic fluctuations around a possible QCP beneath the
superconducting dome. Indications for such QCP have been observed by
other experimental methods \cite{Maiwald2012,Cao2010}. Also, a
similar $T$-linear scattering rate has been reported in hole and
electron doped Ba-based pnictides of the 122 family
\cite{Dai2013a,Lee2015}, yet only for the ND component. The reason for the fact that we also observe $T$-linear scattering for the BD component possibly is related to the difference in substitution method. Hence, it would be interesting to compare our finding to an equivalent study on BaFe$_2$(As$_{1-x}$P$_x$)$_2$. However, to our knowledge there is to date no temperature dependent optical investigation on P-substituted Ba-122 that utilizes the Drude-Lorentz model to decompose the spectra.

In sample P015, the $T$-linear scattering rate at $10 - 300$ K is
only observed for the BD component, while $1/\tau$ of the ND
component is not linear at $T \leq 100$\,K. At $100$\,K, the
localization mode (LOC) appears in the conductivity spectra and
obviously affects the temperature behavior of the ND scattering
rate: $1/\tau$ of the ND term collapses, implying that the carriers
scatter less. The resistivity of this sample shows a flattening at
low temperatures, as seen in the inset of Fig.~\ref{fig:res-mag} (a). Since
$\rho\propto1/(\tau N)$ with $N$ being the carrier density, and
since the BD scattering rate is linear in temperature, the
simultaneous flattening of the $\rho(T)$ curve and collapsing
$1/\tau$ of the ND term can not be consistently understood, if $N$
remains constant. However, in our case $N$ is not constant, because
some electrons get localized. As one can see from Fig.~\ref{fig:fit}
(h), the sum of the spectral weights of the ND and LOC terms remains
basically constant as a function of temperature. The spectral weight
and therefore the number of carriers contributing to the BD term are
also temperature independent. Thus, it is evident that only the
carriers from the ND term are affected by the localization effects.

It is likely that the remaining free carriers of the
ND term scatter less because the contribution of the
electron-electron interactions to the scattering processes becomes
reduced with decreasing the number of free carriers. Alternatively,
the reduction of the ND scattering rate below $100$\,K can be
explained as if the localization mode and the ND term stem from two
different electron bands, which can not be distinguished at elevated
temperatures.

In sample P0165, the spectral weight developments in the FIR region
(Fig.~\ref{fig:fit} (i)) reveal a spectral-weight transfer to lower
frequencies (LOC2 to LOC) with decreasing temperature, reminiscent
of a Drude behavior with collapsing scattering rate. It is important to note,
that the spectral weight of the ND term and of the sum of the
spectral weights of the localization modes stay constant as a
function of temperature. The LOC localization mode in
P0165 (as well as in P015) softens with decreasing temperature,
while the center frequency of the LOC2 mode stays the same at all
temperatures.

\begin{table}[h]
\caption{Parameters of the EuFe$_2$(As$_{1-x}$P$_x$)$_2$ samples
investigated in this work, as extracted from their dc-resistivity
and optical-conductivity measurements. Note, that BD and ND refer to
the broad and narrow Drude components at the \textit{lowest}
measured normal-state temperature ($25$\,K for sample P015; $30$\,K
for samples P0165 and P021). $2\Delta_\text{l}$ and
$2\Delta_\text{s}$ indicate the large and small superconducting gaps
as obtained from the BCS-parametrization fits \cite{Zimmermann1991}
to the measured optical conductivity. The small gap is likely nodal,
thus the value should be considered as an averaged one. $\lambda$ denotes the penetration depth.}
  \begin{tabular}{@{}llll@{}}
    \hline
    Quantity & P015 & P0165 & P021 \\
    \hline
    $\rho @ 300\,$K ($\mu\Omega$cm) & 268  & 507 & 351\\
    RRR  & 2.5 & 5.3 & 4.2\\
        $T_\text{c,on}$ (K)  & 25 & 22 & 27 \\
        $T_\text{c,0}$ (K) & 11.5 & 9 & 25 \\
        BD $\omega_\text{p}$ (cm$^{-1})$ & 8079 & - & 9404 \\
        BD $\tau^{-1}$ (cm$^{-1})$ & 560 & - & 416 \\
        ND $\omega_\text{p}$ (cm$^{-1})$ & 3540 & 2767 & 6235 \\
        ND $\tau^{-1}$ (cm$^{-1})$ & 32 & 14 & 37 \\
        $2\Delta_\text{l}$ (cm$^{-1})$ & 56 ($\pm$ 5) & - &  63 ($\pm$ 3) \\
        $2\Delta_\text{s}$ (cm$^{-1})$ & 12($\pm$ 5) & - & 11($\pm$ 3) \\
        $\lambda$ (nm) & 355 & - & 345 \\
        \hline
  \end{tabular}
  \label{table}
\end{table}

Thus, all three samples demonstrate at least one Drude band with a
directly observable $T$-linear scattering rate. This signals the
importance of electron-electron interactions in Eu-based pnictides
of the 122 family. The localization effects in P015 and P0165
samples strongly affect the optical conductivity and possibly mask
the linearity of $1/\tau$ in other Drude bands. The question remains, why the resistivity of the P015 sample only (Fig. \ref{fig:res-mag} (a)) is affected by the localization effects. This can be explained considering the fact, that the localization effects in P015 appear to be moderate and only affect the carriers when their thermal energy falls below $T \approx 75$ K. In P0165, however, the localization of a part of the carriers persists up to room temperature, while the remaining itinerant carriers follow the observed NFL behavior.

In the MIR and NIR spectral ranges, all three samples display a
spectral-weight transfer from the MIRL interband transition to the
HUND mode upon cooling (Fig.~\ref{fig:fit} (g)). This is a general
observation in iron pnictides and is usually associated to electron
correlations originating from the Hund's-rule coupling between
itinerant and localized electrons \cite{Wang2011,Schafgans2012}.

\section{Optics: superconducting state}
\begin{figure*}[ht]
\includegraphics*[width=\textwidth]{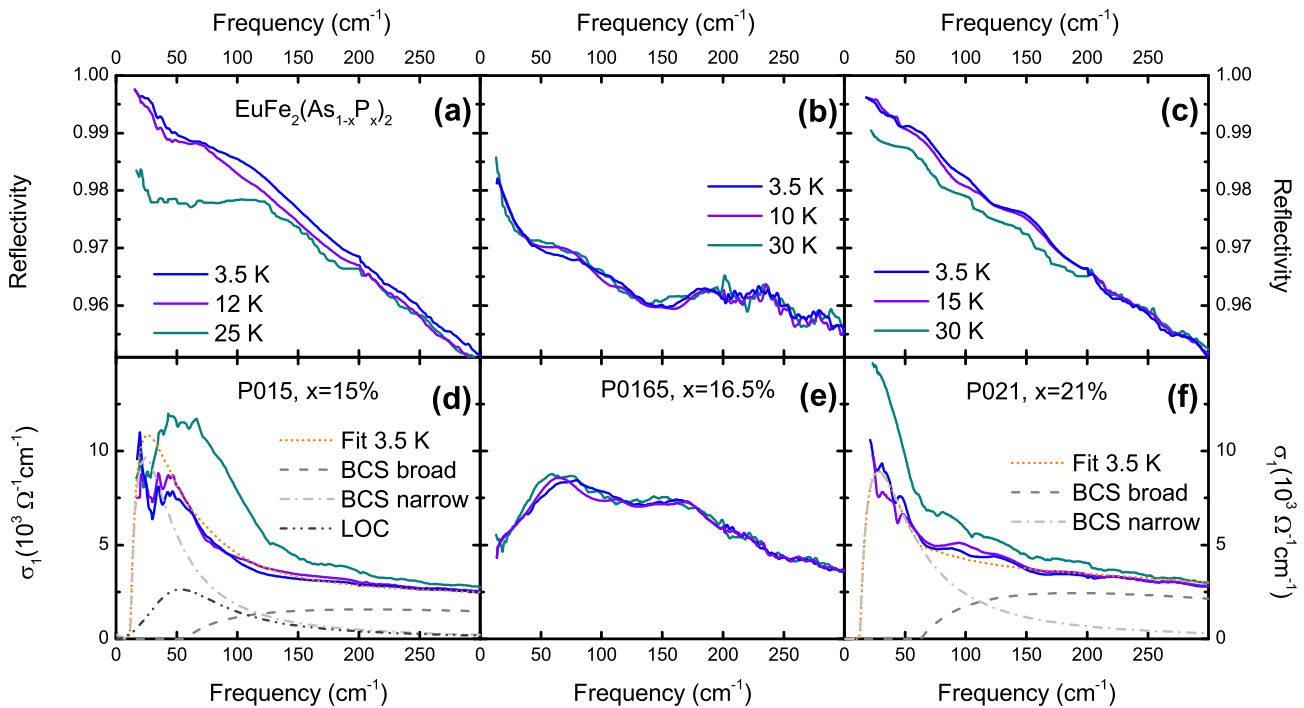}
\caption{Frequency-dependent in-plane reflectivity (panels a-c) and
conductivity (panels d-f) of EuFe$_2$(As$_{1-x}$P$_x$)$_2$ for
$x=15$\%, $x=16.5$\%, and $x=21$\% at one temperature above (cyan)
and two temperatures below (violet, blue) $T_\text{c,on}$. No clear
gap edge can be resolved in any of the spectra. In panels (d) and (f) the resulting BCS fit is depicted as orange dotted lines. The fits contain a broad (dashed grey) and narrow (dash-dotted light grey) BCS term each corresponding to BD and ND, respectively. In panel (f) the fit also containes the suppressed LOC mode in the superconducting state (dark grey dash-double dotted). } \label{fig:sc}
\end{figure*}

In Fig. \ref{fig:sc} (a-c) we show the reflectivity of the
investigated samples for two temperatures below the superconducting transition and one temperature
right above $T_\text{c,on}$. Unlike in the reflectivity spectra of
superconducting EuFe$_2$(As$_{0.82}$P$_{0.18}$)$_2$ previously
published in Ref.\cite{Wu2011}\cite{Note2}, we do not see any sharp upturns in the reflectivity, which
would indicate positions of dirty-limit isotropic superconducting
gaps \cite{Tinkham,Dressel2002}. Neither a full 100\% reflectivity is
observed. Consequently, the conductivity spectra do not show any
sharp gap-like edge either, as seen in Fig.~\ref{fig:sc} (d-f).

For homogenous single-phase superconductors, there are two possible
reasons why no clear signatures of a superconducting gap are
observed. The first option is a strong gap anisotropy, preferably
with nodes: optical transitions of the nodal quasiparticles do not
allow the gap features to manifest themselves in the optical
conductivity \cite{Basov2005}, which is a momentum-averaged
response. The second possibility is superconductivity in the clean
limit. In this case, the scattering rate of the quasiparticles is
much smaller than the superconducting gap, $1/\tau \ll 2\Delta$. As
a consequence, $\sigma_{1}(\omega)$ at frequencies around $2\Delta$
is very close to zero already in the normal state, thus no gap
feature can be detected in the optical-conductivity spectra in the
superconducting state \cite{Kamaras1990}. Below, we argue that
realization of both, the clean-limit superconductivity and the gap
anisotropy, are likely in EuFe$_2$(As$_{1-x}$P$_x$)$_2$. Let us note
that our optical measurements, performed above and below $T_{N}$,
show that the ordering of the Eu magnetic moments at $T_{N}$ does not
affect the optical conductivity, in agreement with previous optical studies on the parent compound \cite{Wu2009,Zapf2015,Moon2010}. Thus, influence of the
Eu-moments ordering on the optical-gap feature is very unlikely.

Although a full multiband Eliashberg analysis is the most adequate
approach for fitting the optical conductivity of iron pnictides in
the superconducting state \cite{Charnukha2011}, application of this
method requires advanced theoretical methods and goes beyond our
capability. During the last years, it has been shown by different
groups \cite{Wu2010b,Kim2010,Maksimov2011a,Dai2013b} that applying a
relatively simple model \cite{Zimmermann1991}, which parameterizes
the optical response of a BCS superconductor, is often sufficient in
order to get a rough estimate for the magnitude of the
superconducting gap(s) in pnictides.

Even though we do not observe any clear gap features we can use this model to fit the spectrum of sample P021 in the
superconducting state with the constraint, that the plasma
frequencies and scattering rates of the BCS terms, as well as all
parameters of the Lorentz contributions, stay roughly the same as in
the normal state just above the superconducting transition.
In this sample, the optical response of free
carriers is not masked by the appearance of localization modes, thus the
analysis is relatively straightforward. Two BCS terms, each
replacing a normal-state Drude term, were utilized in this fit, as displayed Fig.\ref{fig:sc} (f). We
obtained the gap values of $2\Delta_{l} = 63 \pm 3$\,cm$^{-1}$ and
$2\Delta_{s} = 11 \pm 3$\,cm$^{-1}$ for the two BCS terms, which
replaced the broad and narrow Drude terms, respectively. 

For the larger gap, we obtain $2\Delta/k_\text{B}T_\text{c}\approx
3.5$, which is close to the BCS value. For the small gap, this ratio
is below $1$. This indicates that the isotropic gap is likely not a
relevant approach and points towards anisotropy of the smaller gap
(a moderate anisotropy of the larger gap is also possible). Since
the BCS term with the large / small gap in the superconducting state
corresponds to the broad / narrow Drude term in the normal state, we
can interpret our data in terms of an (almost) isotropic gap on the
hole pockets and a highly anisotropic (or nodal) gap on the electron
pockets of the Fermi surface. This is in agreement with multiple
reports on the sister system BaFe$_2$(As$_{1-x}$P$_x$)$_2$
\cite{Shibauchi2014,Hashimoto2010,Nakai2010,Ishikado2011,Yamashita2011},
as well as with theoretical predictions for P-substituted iron
pnictides \cite{Thomale2011,Chubukov2009}.

\begin{figure}[ht]
\includegraphics[width=\linewidth]{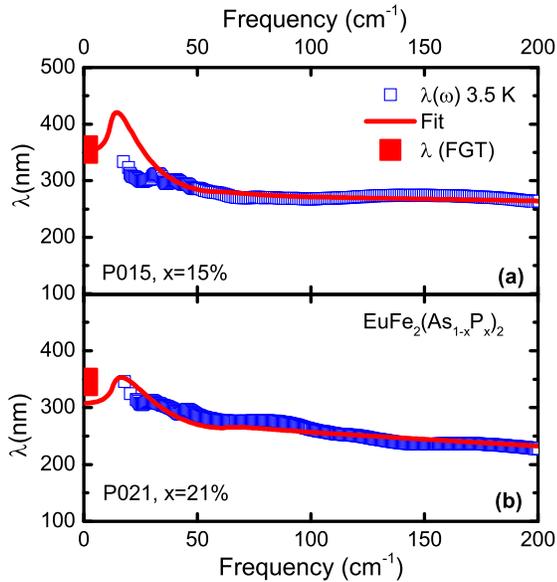}
\caption{Frequency-dependent penetration depth $\lambda(\omega)$ of EuFe$_2$(As$_{1-x}$P$_x$)$_2$ at $3.5$\,K for
(a) $x=15$\% and (b) $x=21$\%. Blue open symbols are data points, while red lines represent the two gap BCS fits as described in the text. Large red symbols at zero frequency denote the values of the penetration depth obtained by use of the FGT sum rule.} \label{fig:lamb}
\end{figure}

To examine whether our samples are in the clean or dirty limit, it
is useful to calculate the frequency-dependent penetration
depth, $\lambda(\omega)=c/[4\pi\omega\sigma_2(\omega)]^{1/2}$. The
penetration depth for sample P021 at $3.5$\,K is shown in Fig. \ref{fig:lamb} (b).
$\lambda(\omega)$ is basically flat, with a small gradual increase
towards low frequencies. This is indicative for clean-limit
superconductivity \cite{Lobo2015} and consistent with reports on
BaFe$_2$(As$_{1-x}$P$_x$)$_2$ \cite{Moon2012}. The model
$\lambda(\omega)$, computed from our BCS fit, describes the
experimental points fairly accurately, see Fig. 5.

Alternatively, one can obtain the penetration depth from the
temperature evolution of the $\sigma_1(\omega)$ spectra. The
occurrence of the superfluid condensate implies a transfer of the
spectral weight from finite frequencies to a $\delta(\omega=0)$
function in $\sigma_{1}(\omega)$, representative of the infinite dc
conductivity in the superconducting state. As the measured optical $\sigma_{1}(\omega)$ has no access to zero
frequency, the value of its integral drops when the superfluid
forms. The difference between the spectral weights in the normal and
superconducting states is directly related to the superfluid density
$\rho_\text{s}$ (the Ferell-Glover-Tinkham (FGT) sum rule
\cite{Tinkham,Dressel2002}):
\begin{equation}
\rho_\text{p,s}=8\int_0^{\omega_\text{c}}\left[\sigma_1(\omega,
30\,K)-\sigma_1(\omega,3.5\,K)\right] d\omega
        \label{eq:FGT}
\end{equation}
and allows calculation of the penetration depth \textit{via}
$\lambda=c/(\rho_\text{s})^{1/2}$. The cutoff frequency
$\omega_\text{c}$ needs to be chosen such, that all the spectral
weight transferred to the $\delta(\omega=0)$ peak is captured. We
find $\omega_\text{c}=500\,\text{cm}^{-1}$ to be a reasonable value
for this purpose. Calculations using Eqn.~\ref{eq:FGT} provide a
penetration depth of $345$\,nm, which agrees very well with the
value obtained from $\sigma_2(\omega)$, see Fig. \ref{fig:lamb}, and thus, shows
the consistency of our analysis.

Let us now turn to samples P015 and P0165. In these specimens, the
behavior of the reflectivity [Fig.~\ref{fig:sc} (a, b)] and the
optical conductivity [Fig.~\ref{fig:sc} (d, e)] in the
superconducting state is different as compared to sample P021,
because of the appearance of the localization mode(s) at low
frequencies. While in P0165 there is no clear difference between the
optical responses in the normal and superconducting state, in P015
the LOC mode appears to be suppressed and shifted towards lower
frequencies below $T_\text{c,on}$. Again, a clear gap edge is not
observed.

In order to perform a meaningful analysis of the penetration depth
in P015, it is necessary to subtract the contribution of the
localization mode from $\sigma_2(\omega)$ before calculating
$\lambda(\omega)$. When we do so, we obtain the low-frequency
$\lambda(\omega)$ of P015 to be rather similar to $\lambda(\omega)$ of P021,
as one can see from Fig. \ref{fig:lamb}. The penetration depth, calculated using
the FGT sum rule yields now $\lambda=355$\,nm, slightly larger than in
P021, but in good agreement with the frequency-dependent $\lambda$,
see Fig. \ref{fig:lamb} (a) .

For P015, the fit with the BCS terms in the superconducting state
requires a spectral-weight transfer from the LOC mode to the ND
component as temperature decreases and provides $2\Delta= 56 (\pm
5)\,\text{cm}^{-1}$ and $2\Delta= 12 (\pm 5)\,\text{cm}^{-1}$, which
is very similar to gap values in P021 (see Fig. \ref{fig:sc} (d)). Also similarly to P021, the
penetration depth, obtained from the BCS fits, describes the
experimental $\lambda(\omega)$ very well, as shown in Fig. 5.

Thus, we can conclude that the optical responses of itinerant
carriers in samples P015 and P021 are very similar to each other;
the differences in the doping level do not affect significantly the
values of the superconducting gaps. The only major difference
between these two samples, seen in the optical response, is the
localization mode, which is present in P015 and absent in P021. In
the sample with an intermediate doping level, P0165, the absorption
due to localization effects is even stronger than in P015. This
absorption (modeled above with two Lorentzian terms) completely
prevents the observation of any superconductivity-induced changes in
the optical spectra of P0165, even though bulk superconductivity was
confirmed in this sample by different magnetization measurements, as
discussed in Sec. 3. We pointed out above, the
non-systematic behavior of the FIR bands in our
EuFe$_2$(As$_{1-x}$P$_x$)$_2$ samples as a function $x$ is evident
for extrinsic character of the FIR absorption bands. If the density
of localized carriers is not very large (the case of sample P015),
the optical response remains to be primarily dominated by itinerant
carriers. In the case of very impure/disordered samples (sample
P0165), the optical response becomes fully determined by localized
carriers. Looking on the entire set of our experimental data, we can
not exclude that sample P0165 has at least two phases, only one of
which is superconducting. The same might be true for sample P015, but
the fraction of the superconducting phase is surely much larger in
this sample.

\section{Conclusions}
We performed a detailed study of broadband optical response of
single-crystalline superconducting EuFe$_2$(As$_{1-x}$P$_x$)$_2$
with $x=15$\%, $16.5$\%, and $21$\%. The former two samples show
moderate ($x=15$\%) to strong ($x=16.5$\%) impurity-localization
effects, while the later ($x=21$\%) appears to be very pure. All of
the samples exhibit superconductivity with onsets at comparable
temperatures, basically independent of the doping level and
appearance of the localization effects. The response of the
itinerant carriers in the normal state can be generally best
described by two Drude components, which might be related to the
hole and electron pockets of the Fermi surface, similarly to other
pnictides \cite{Charnukha2014,Wu2010b}. We were able to uncover a
linear in temperature scattering rate for at least one of the two
Drude terms in each of the samples. In the pure $21$\%-doped sample,
the linearity is apparent for both bands. In two other samples, the
linearity is (partly) masked by strong FIR modes due to the
localization effects. This linearity (as well as the linear
temperature dependence of dc resistivity) points towards a
non-Fermi-liquid state in EuFe$_2$(As$_{1-x}$P$_x$)$_2$, which leads to speculations wether 
a QCP exists beneath the superconducting dome, as it is evident in the sister compound BaFe$_2$(As$_{1-x}$P$_x$)$_2$.

Analysis of the frequency-dependent optical spectra in the
superconducting state allows us to conclude that
EuFe$_2$(As$_{1-x}$P$_x$)$_2$ is likely a clean-limit superconductor
with two gaps: a smaller anisotropic (or nodal) gap on the electron
portion of the Fermi surface and a larger isotopic (or slightly
anisotropic) gap on the hole pockets. This is in agreement with
theoretical predictions as well as with experimental observations
made in other P-substituted 122 iron pnictides.

\section{Acknowledgements}
We are grateful to G. Untereiner, C. Kamella and E. Rose for
technical support. We thank U. S. Pracht, S. Jiang and D. Wu for
helpful discussions. This project was funded by the DFG SPP 1458.


\end{document}